\documentclass{emulateapj}
\usepackage{graphicx}    
\usepackage{subfigure}
\usepackage{color}
\newcommand{\Msun}{h^{-1} M_{\odot}}
\submitted{}

\begin{document}
\title{Galaxy Mergers and Dark Matter Halo Mergers in LCDM: Mass, Redshift, and Mass-Ratio Dependence}
\author{Kyle R. Stewart\altaffilmark{1}, James S. Bullock\altaffilmark{1},
Elizabeth J. Barton\altaffilmark{1}, Risa H. Wechsler\altaffilmark{2}}

\altaffiltext{1}{Center for Cosmology, Department of Physics and Astronomy, The University of California at Irvine, Irvine, CA, 92697, USA}
\altaffiltext{2}{Kavli Institute for Particle Astrophysics \&
  Cosmology, Department of Physics, and SLAC National Accelerator
  Laboratory, Stanford University, Stanford, CA 94305, USA}

\begin{abstract} {
We employ a high-resolution LCDM N-body simulation to present merger rate predictions
for dark matter halos and investigate how common
merger-related observables for galaxies---such as close pair counts, starburst counts,
and the morphologically disturbed fraction---likely scale with luminosity,
stellar mass, merger mass ratio, and redshift from $z=0$ to $z=4$.
We investigate both rate at which subhalos first enter
the virial radius of a larger halo (the ``infall rate''), and the rate
at which subhalos become destroyed, losing $90\%$ of the
mass they had at infall (the ``destruction rate'').
For both merger rate definitions, we provide a simple `universal'
fitting formula that describes our
derived merger rates for dark matter halos a function of dark halo mass,
merger mass ratio, and redshift, and go on to predict galaxy merger rates
using number density-matching to associate halos with galaxies.  For example,
we find that the instantaneous (destruction) merger rate of $ m/M > 0.3$ mass ratio events
into typical $L \gtrsim f\, L_*$ galaxies follows the simple relation
$dN/dt \simeq 0.03 (1+f) \, {\rm Gyr}^{-1} \, (1+z)^{2.1}$.  Despite
the rapid increase in merger rate with redshift, only a small fraction
of $>0.4 L_*$ high-redshift galaxies ($\sim 3\%$ at $z=2$) should have
experienced a major merger ($m/M >0.3$) in the very recent past ($ t<
100$ Myr).  This suggests that short-lived, merger-induced bursts of
star formation should not contribute significantly to the global star
formation rate at early times, in agreement with several observational
indications.  In contrast, a fairly high fraction ($\sim 20\%$) of
those $z = 2$ galaxies should have experienced a morphologically
transformative merger within a virial dynamical time ($\sim 500$ Myr at $z=2$).  We compare our
results to observational merger rate estimates from both morphological
indicators and pair-fraction based determinations between $z=0-2$ and
show that they are consistent with our predictions.
However, we emphasize that great care must be made in these comparisons because
the predicted observables depend very sensitively on galaxy luminosity,
redshift, overall mass ratio, and uncertain relaxation timescales for merger remnants.
We show that the {\em majority} of bright galaxies at $z = 3$
should have undergone a major merger ($>0.3$) in the previous
$700$ Myr and conclude that mergers almost certainly play an important
role in delivering baryons and influencing the kinematic properties of
Lyman Break Galaxies (LBGs).}
\end{abstract}


\section{Introduction}
\label{Introduction}

In the current theory of hierarchical structure formation (LCDM), dark
matter halos and the galaxies within them are assembled from the
continuous accretion of smaller objects
\citep{Peebles82,Blumenthal84,Davis85,Wechsler02,FakhouriMa08,Stewart07,Cole08,NeisteinDekel08,Wetzel08}.
It is well-established that galaxy and halo mergers should be more
common at high redshift
\citep[e.g.][]{Governato99,Carlberg00,Gottlober01,Patton02,Berrier06,Lin08,Wetzel08},
but the precise evolution is expected to depend on details of the
mergers considered.  Moreover, it is unclear how these mergers
manifest themselves in the observed properties of high-$z$ galaxies
and what role they play in setting the properties of galaxies in the
local universe.  Interestingly, there are indications that the
familiar bimodality of galaxies as disks versus spheroids at $z=0$ might be
replaced by a categorization of disk-like versus merger-like at higher
redshift \citep{Schreiber06,Genzel06,Law07,Kriek08,Melbourne08,
  Shapiro08, Wright08}, although this shift in the dichotomy of
galaxy morphologies is by no means robust and requires further study.
In this paper, we use N-body simulations to provide
robust predictions and simple fitting functions for dark matter halo
merger rates and merger fractions as a function of redshift, mass, and
mass ratio.  We use our predictions to address two observable
consequences of galaxy mergers---merger-driven starbursts and
morphological disturbances---and investigate their evolution with
redshift.

The tidal interactions inherent in galaxy mergers produce
concentrations of gas in the remnant centers.
For major mergers ($m/M \gtrsim 0.3$), models predict that this effect results
in a significant burst of increased star formation rate (SFR) compared to the central galaxy's
past star formation history \citep[e.g.][]{MihosHernquist96,cox07}.
It is also likely to enable supermassive black hole
growth and the fueling of AGN  \citep[e.g.,][]{Heckman86, Springel05b, Somerville08}.
\cite{cox07} used Smooth Particle Hydrodynamical (SPH) simulations to
show that   the timescale over which merger-induced  starbursts are active
depends sensitively on the treatment of poorly-understood feedback and ISM physics; they demonstrate
that future observational constraints on this timescale
may provide a means to constrain feedback models  \citep[][and references therein]{Barton07}.
Historically, SPH simulations have treated star forming gas as isothermal, and this treatment
results in starburst timescales in major mergers that are quite short-lived, with
$t_* \sim 100$ Myr \citep[e.g.][]{MihosHernquist96, cox07}.
Recently, it has become popular in SPH simulations to impose a stiff equation of
state for star-forming gas in order to mimic the effects of a multi-phase ISM and to
suppress star formation and disk fragmentation \citep{Yepes97, SpringelHernquist03, Governato07}.
\cite{cox07} showed that a stiff equation of state of this kind
significantly lengthens the timescale for starburst activity in major mergers to $t_* \sim 500$ Myr.
Below we investigate the evolution of merger fractions with $100$ Myr and $500$ Myr as a first-order
means of addressing the differences between merger-induced starburst fractions in different feedback schemes.

A second observationally-relevant consequence of  mergers is morphological disturbance.
Very large mergers, especially those with moderately low gas fractions, likely play a role in
transforming late type disk galaxies into ellipticals
\citep[e.g.,][]{ToomreToomre, BarnesHernquist96,Robertson06a,Robertson06b,Burkert08}.
If gas fractions are high in major mergers (as expected at high redshift) then they
may play a role in building early disks
\citep{Robertson06a,Hopkins08g, Hopkins08i, RobertsonBullock08}.  More common are
moderate-size ($m/M > 0.1$) dark matter halo mergers \citep{Stewart07},
which can produce morphological signatures like
disk flaring, disk thickening, and ring and bar-like structures in disk galaxies
\citep{BarnesHernquist96,k07,Younger07,VillalobosHelmi08,Purcell08b} as well as tidal features seen in massive elliptical
galaxies \citep{Feldmann08}.

Below we explore two possibilities for the evolution of the morphological relaxation time with redshift.
First, we explore a case where the remnant  relaxation time scales with redshift, approximated
by the dark matter halo dynamical time ($\tau \propto (1+z)^{-\alpha}$, $\alpha \simeq 1.1-1.5$; see below),
and second we investigate the possibility that relaxation times remain constant
with redshift at $\tau \simeq 500$ Myr.  The latter timescale is motivated by the results of \cite{lotz08b}
who studied outputs from SPH merger simulations of $z=0$ galaxies in great detail
\citep[see][for additional descriptions of these simulations and their analysis]{cox_jonsson06,jonsson06,rocha08,cox07}.
 These choices bracket reasonable expectations and allow us to
provide first-order estimates for the evolution in the morphologically disturbed fraction with
redshift.  More simulation work is needed to determine how the relaxation times
of galaxy mergers should evolve with redshift, including an allowance for the evolution in
approach speeds, galaxy densities, and orbital parameters (if any).

Though not discussed in detail here, a third consequence of mergers is
the direct, cumulative deposition of cold baryons (gas and stars) into galaxies.  For this question, one is interested
in the full merger history of individual objects, rather than the instantaneous merger rate or recent merger fraction.
Specifically, one may ask about the total mass that has been deposited by major mergers over a galaxy's history.
We focus on this issue in \cite{Stewart09b}.

In what follows we use a high-resolution dissipationless cosmological LCDM N-body
simulation to investigate the merger rates and integrated merger fractions
of galaxy dark matter halos of mass $M = 10^{11} - 10^{13} \Msun$
from redshift $z=0$ to $4$.  We adopt the simple technique of monotonic
abundance-matching in order to associate dark matter halos with galaxies
of a given luminosity or stellar mass
\citep[e.g.,][]{Kravtsov04a, Conroy06, Berrier06, ConroyWechsler08},
and make predictions for the evolution of the galaxy merger rate with redshift.

The outline of this paper is as follows.  In \S \ref{Simulation} we
discuss the numerical simulation used and the method of merger tree
construction, while we present merger statistics for dark matter halos
in \S \ref{DM}.
In \S \ref{Methods} we discuss the method of assigning
galaxies to dark matter halos both as a function of stellar mass, and
alternatively as a function of galaxy luminosity (compared to $L_*(z)$).
In \S \ref{Results} we present our principle results,
which characterize the merger rate of galaxies as a function of redshift, with
comparison to observed
properties of bright galaxies.
We summarize our main conclusions in \S \ref{Conclusion}.

\section{Simulation}
\label{Simulation}

We use a simulation containing $512^{3}$ particles, each with mass
$m_p = 3.16 \times 10^8 \Msun$, evolved within a comoving cubic volume
of $80 h^{-1}$ Mpc on a side using the Adaptive Refinement Tree (ART)
$N$-body code \citep{Kravtsov97,Kravtsov04a}.  The simulation uses a
flat, $\Lambda$CDM cosmology with parameters
$\Omega_{M}=1-\Omega_{\Lambda}=0.3$, $h=0.7$, and $\sigma_{8}=0.9$.
The simulation root computational grid consists of $512^3$ cells,
which are adaptively refined to a maximum of eight levels, resulting
in a peak spatial resolution of $1.2 h^{-1}$ kpc (comoving).  Here we
give a brief overview of the simulation and methods used to construct
the merger trees.  They have been discussed elsewhere in greater
detail \citep{allgood06,wechsler06,Stewart07} and we refer the reader
to those papers for a more complete discussion.

Field dark matter halos and subhalos are identified using a variant of
the bound density maxima algorithm \citep{Klypin99}.  A \emph{subhalo}
is defined as a dark matter halo whose center is positioned within the
virial radius of a more massive halo.  Conversely, a \emph{field halo}
is a dark matter halo that does not lie within the virial radius of a
larger halo.  The virial radius $R$ and mass $M$ are defined such that
the average mass density within $R$ is equal to $\Delta_{\rm vir}$
($\simeq 337$ at $z=0$) times the mean density of the universe at that
redshift.  Our halo catalogs are
complete to a minimum halo mass of $M =10^{10} \Msun$, and our halo
sample includes, for example, $\sim 15,000(10,000)$ and $2,000(500)$
field halos at $z=0(3)$ in the mass bins $10^{11-12}$ and $10^{12-13}
\Msun$, respectively.

We use the same merger trees described in \cite{Stewart07},
constructed using the techniques described in \cite{Wechsler02} and
\cite{wechsler06}.  Our algorithm uses 48 stored timesteps that are
approximately equally spaced in expansion factor between $a =
(1+z)^{-1} = 1.0$ and $a = 0.0443$.  We use standard terminologies for
progenitor and descendant.  Any halo at any timestep may have any
number of progenitors, but a halo may have a single descendant ---
defined to be the halo in the next timestep that contains the majority
of this halo's mass.  The term {\em main progenitor} is used to
reference the most massive progenitor of a given halo, tracked back in
time.

Throughout this work we present results in terms of the {\em merger
  ratio} of an infalling object, $m/M$, where we always define $m$ as
the mass of the smaller object just \emph{prior} to the merger and $M$
is the mass {\em main progenitor} of the larger object at the same
epoch.  Specifically, $M$ in the ratio does not incorporate the mass
$m$ and therefore $m/M$ has a maximum value of $1.0$.  Except when
explicitly stated otherwise, we always use dark matter halo masses
to define the merger ratio of any given merger event, and we always define
the merger ratio as the mass ratio just before the smaller halo falls
into the virial radius of the larger one.  Because there is
not a simple linear relation between halo mass and galaxy stellar (or baryonic)
mass, this is an important distinction.  For example, our major mergers,
defined by halo mass ratios, may not always correspond to major
galaxy mergers as defined by stellar or baryonic mass ratios
\citep[see e.g.][]{Stewart09conf}.

In what follows we investigate two types of mergers.  The first and
most robust of our predicted rates is the {\em infall rate}: the rate
at which infalling halos become subhalos, as they first fall within
the virial radius of the main progenitor.  These are the results we
present in Section 3, which describes our `universal' merger rate
function for dark matter halos.  The second rate is aimed more closely
at confronting observations and is associated with central mergers
between galaxies themselves.  Specifically we define the {\em
destruction rate} by counting instances when each infalling subhalo
loses $90\%$ of the mass it had prior to entering the virial radius of
the larger halo.~\footnote{Subhalo masses are defined to be the mass
within a truncation radius $R_t$, which is set to be the minimum of
the virial radius and the radius where the subhalo density profile
begins to encounter the background halo density.}  We are unable to
measure central crossings directly because the time resolution in our
snapshot outputs (typically $\Delta t \simeq 250$ Myr) is comparable
to a galaxy-galaxy crossing time at the centers of halos,
however, for mergers with mass ratios $>1/3 (1/10)$, subhalo destruction
typically takes place $\sim2 (3)$ Gyr after infall to the virial radius.
Based on simulations of galaxy mergers,
this definition leads to subhalo ''destruction''
sometime after its first pericenter (and likely after second pericenter),
but probably before final coalescence \citep{BoylanKolchin08}.
Note that this second rate (destruction) is
more uncertain than the first (infall) because, in principle, the orbital
evolution of infalling galaxies will depend upon the baryonic
composition of both the primary and secondary objects.  Fortunately,
as presented in detail below (see Table 1) for the relatively high
mass-ratio merger events we consider, the merger rates (and their
evolution with redshift) do not depend strongly on whether we define a
merger to occur at halo infall or at this central mass-loss epoch.

\begin{figure*}[t!]
  \plotone{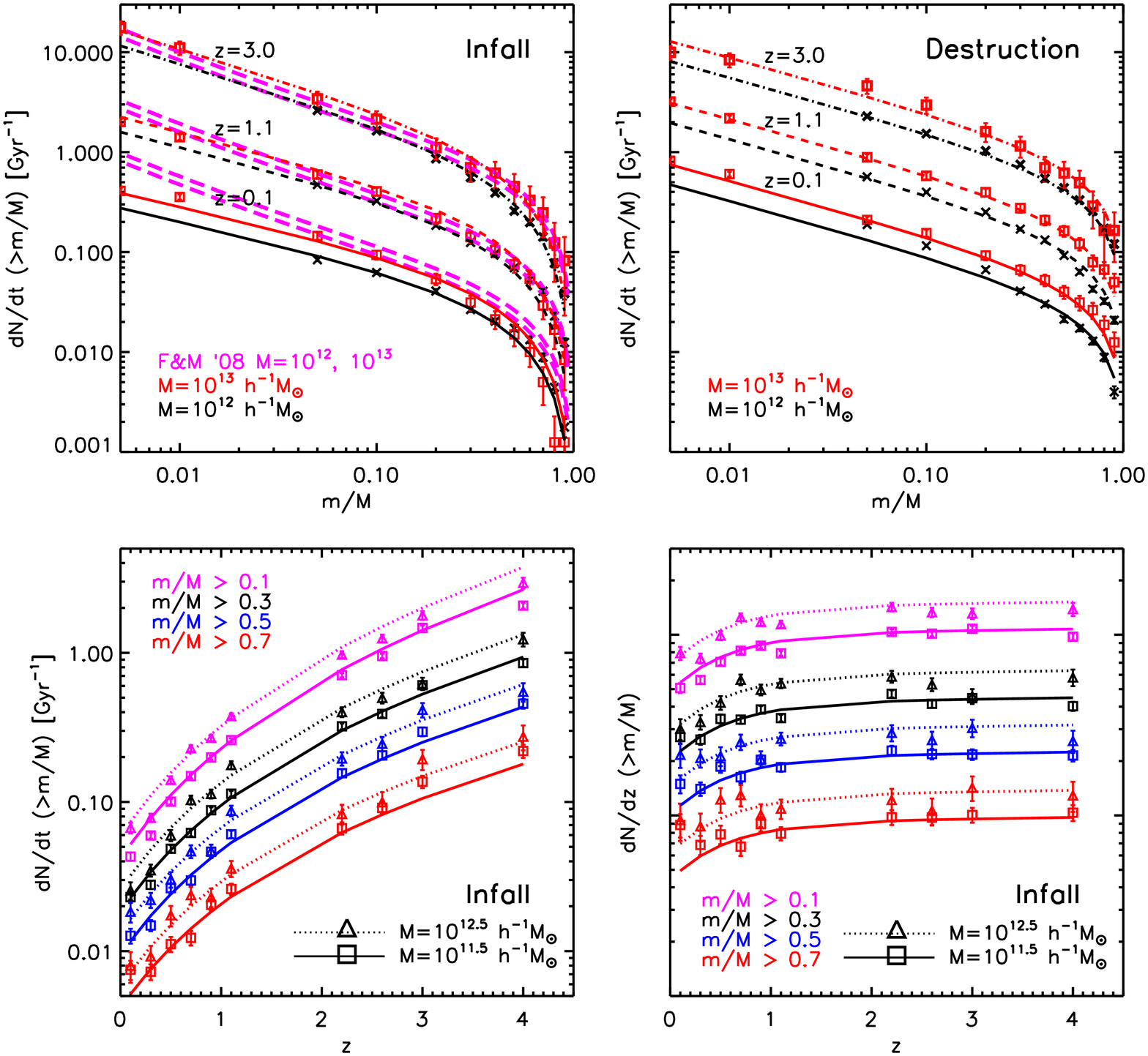}
  \caption{Dark matter halo infall and destruction rates (see \S \ref{Simulation})
as a function of mass, merger-mass ratio, and redshift.
Host halo mass bins span $\Delta \log_{10} M = 0.5$.
\emph{Top Left:} Infall rate per Gyr as a function
of merger mass ratio.  The dashed (pink) lines are a comparison
to the results of \cite{FakhouriMa08} for $M=10^{12} \Msun$ (lower) and
$M=10^{13} \Msun$ (upper) halos.
\emph{Top Right:} Identical to top left, but for the destruction rate of halos
instead of the infall rate.
\emph{Bottom Left:} Infall rate per Gyr as a function of redshift.
\emph{Bottom Right:} Infall rate per unit redshift, as a function of redshift.
In the top panels, black and red lines correspond to different
host halo masses ($10^{12}$ and $10^{13} \Msun$),
while in the bottom panels, magenta, black, blue and red lines correspond to different merger
ratios ($m/M > 0.1, 0.3, 0.5, 0.7$).
In all panels, the mass ratio, $m/M$ is defined prior to the infall
of the smaller system.  Error bars are Poissonian based on
the number of host halos and the number of mergers.
Horizontal error bars on the bottom figures
have been omitted for clarity, but are identical to those in Figure 2.}
\label{fig:dm_rate}
\end{figure*}

\begin{table*}[thb!]
\begin{center}
\caption{Merger Rate Fitting Function Parameters for Equations 3.1,3.2, and 3.3.}
  \begin{tabular}{ | l | l | l | l | l |}    \hline
  {\bf Dark Matter Halos}: 	& $A(z,M)$\tablenotemark{$\dagger$} 			& $c$		& $d$ 			\\
  ($M = 10^{11.0-13.5} \Msun$) & & & \\\hline
  dN/dt:  (INFALL) simple fit & $0.020 \, (1+z)^{2.3} \, M_{12}^{0.15}$ 		& $0.50$	& $1.30$ 		 \\ \hline
  dN/dt:  (INFALL) complex fit & $0.020 \, (1+z)^{2.3} \, M_{12}^{0.15}$  		& $0.4+.05z$	& $1.30$ 		 \\ \hline
  dN/dz:  (INFALL)    		& $0.27 \, (d\delta_c/dz)^2  \, M_{12}^{0.15}$ 		& $0.50$	& $1.30$ 		 \\ \hline
  dN/dt:  (DESTROYED)   	& $0.022 \, (1+z)^{2.2} \, M_{12}^{0.2}$  		& $0.54$	& $0.72$ 		\\ \hline
  dN/dz:  (DESTROYED)    	& $0.32 \, (d\delta_c/dz) \, M_{12}^{0.2} $ 		& $0.54$	& $0.72$ 		 \\ \hline
  \hline
  {\bf Galaxy Luminosity Cuts}: 			  	& $A(z,f)$\tablenotemark{$\dagger$}	  & $c$		&$d$ 	\\
  ($L > f L_*$, $0.1<f<1.0$) & & & \\ \hline
  dN/dt:		    					& $0.02 \, (1 + f) \, (1+z)^{2.1}$   	  & $0.54$	&$0.72$	\\ \hline
  Merger Fraction in past $T$ Gyr: 				& $0.02 \, T \, (1 + f) \, (1+z)^{2.0}$   & $0.54$	&$0.72$	 \\
  (Frac $< 0.6$, $T < 4$) & & & \\ \hline
  \hline
  {\bf Galaxy Stellar Mass Ranges}: 			 & $A(z)$\tablenotemark{$\dagger$}	& $c_*$\tablenotemark{$\ddagger$}	& $d_*$\tablenotemark{$\ddagger$} \\
  ($F(x) = F(m_*/M_*)$) & & & \\ \hline
  dN/dt ($10^{10.0} M_{\odot}<M_*<10^{10.5}M_{\odot}$):	& $0.015 \, e^{1.0 z}$ \hspace{3em}  		& $0.30 \,  \, $	& $1.1-0.2z$		\\ \hline		
  dN/dt ($10^{10.5}M_{\odot}<M_*<10^{11.0}M_{\odot}$):	& $0.035 \, e^{0.7 z}$ \hspace{3em}  		& $0.25 \,  \, $	& $1.1-0.2z$		\\ \hline		
  dN/dt ($10^{11.0}M_{\odot}<M_*$):			& $0.070 \, e^{1.0 z}$ \hspace{3em}  		& $0.20 \,  \, $	& $1.0-0.3z$		\\ \hline		
  \end{tabular} \\
  \end{center}
  \tablenotemark{$\dagger$}When not dimensionless, units are Gyr$^{-1}$. \\
  \tablenotemark{$\ddagger$}Mass-ratio variable for galaxy stellar mass merger rates are identified with a \emph{stellar mass} ratio, $r = m_*/M_*$. \\
  \label{B_FitParams}
 \end{table*}

\begin{table*}
\begin{center}
\caption{Dark Matter Halo Mass--Luminosity Relationship by Number Density Matching.}
  \begin{tabular}{| l | l | cc | cc | cc | c |}
    \tableline
    \tableline
    $z$ & Source \& Rest-Frame Band  & \multicolumn{2}{c |}{$>0.1L_*$} & \multicolumn{2}{c |}{$>0.4L_*$} & \multicolumn{2}{c |}{$> L_*$} &  $\tau(z)$\\
    \cline{3-4} \cline{5-6} \cline{7-8}
      &             	& \multicolumn{2}{l |} {$n_g$ \tablenotemark{$\dagger$} \, \,  \, $M_{DM}$ \tablenotemark{$\ddagger$}}
      				& \multicolumn{2}{l |} {$n_g$ \tablenotemark{$\dagger$} \, \,  \, $M_{DM}$ \tablenotemark{$\ddagger$}}
      				& \multicolumn{2}{l |} {$n_g$	\tablenotemark{$\dagger$} \, \,  \, $M_{DM}$  \tablenotemark{$\ddagger$}}
				&  [Gyr] \\ \hline
  $0.1$ & SDSS (r$^{0.1}$-band) \tablenotemark1    	& 29 & $10^{11.2}$ 	& 10	& $10^{11.7}$ & 3.2 & $10^{12.3}$ & 1.79 \\ 
  $0.3$ & COMBO-17/DEEP2 (B-band) \tablenotemark2 	& 20 & $10^{11.4}$ 	& 5.8 	& $10^{12.0}$ & 1.5 & $10^{12.6}$ & 1.50 \\                             
  $0.5$ & COMBO-17/DEEP2 (B-band) \tablenotemark2	& 24 & $10^{11.3} $	& 7.0 	& $10^{11.9}$ & 1.8 & $10^{12.5}$ & 1.28  \\	
  $0.7$ & COMBO-17/DEEP2 (B-band) \tablenotemark2	& 20 & $10^{11.4} $	& 5.8 	& $10^{12.0}$ & 1.5 & $10^{12.6}$ & 1.09 \\  
  $0.9$ & COMBO-17/DEEP2 (B-band) \tablenotemark2	& 25 & $10^{11.3} $	& 7.3  	& $10^{11.9}$ & 1.9 & $10^{12.5}$ & 0.95 \\ 
  $1.1$ & COMBO-17/DEEP2 (B-band) \tablenotemark2	& 19 & $10^{11.4} $	& 5.5 	& $10^{12.0}$ & 1.4 & $10^{12.5}$ & 0.81 \\ 
  $2.2$ & Keck Deep Fields (UV) \tablenotemark3   	& 20 & $10^{11.3}$   	& 6.4 	& $10^{11.7}$ & 1.8 & $10^{12.2}$ & 0.49 \\
  $\sim 2.6$ & Extrapolation  (UV)			& $\sim 18$ & $10^{11.3}$ & $\sim 5.2$ & $10^{11.8}$ & $\sim 1.2$ & $10^{12.3}$ & 0.40 \\
  $3$ &Keck Deep Fields (UV)  \tablenotemark3      	& 15 & $10^{11.2}$ 	& 3.8 	& $10^{11.7}$ & 0.90 & $10^{12.2}$ & 0.32 \\
  $3$ &Lyman Break Galaxies (V-band) \tablenotemark4   & NA 	& NA  		& 5.0 	& $10^{11.7}$ & 0.82 & $10^{12.2}$ & 0.32 \\
  $4$ & HUDF + HST ACS Fields (UV) \tablenotemark5 	& 18	 & $10^{11.0}$ 	& 3.2 	& $10^{11.6}$ & 0.61 & $10^{12.0}$ & 0.25 \\
  \tableline
  \end{tabular}
  \label{LstarTable}
   \tablenotetext{$\dagger$}{$10^{-3}$ $h^{3}$ Mpc$^{-3}$}
   \tablenotetext{$\ddagger$}{$\Msun$}
   \tablenotetext1{\cite{Blanton03}}
   \tablenotetext2{\cite{Faber07} with DEEP2 optimal weights.}
   \tablenotetext3{\cite{Sawicki06}}
   \tablenotetext4{\cite{Shapley01} -- note that rest-frame V number densities match well with $> 0.4 L_*$ and $> L_*$ values  in rest-UV at $z=3$.}
   \tablenotetext5{\cite{Bouwens07}}
\end{center}
\end{table*}

\section{Dark Matter Halo Merger Rates}
\label{DM}
We begin by investigating infall and destruction merger rates as a
function of mass, merger ratio, and redshift.  Merger rates are
shown for several of these choices in the four panels of Figure
\ref{fig:dm_rate}.  The upper panels show merger rates per unit time
for $>m/M$ mass ratio objects falling into host halos of mass
$10^{12}$ (black lines and crosses) and $10^{13} \Msun$ (red lines and squares)
at three different redshifts: $z=0$ (solid), $z=2$ (dashed), and $z=3$
(dot-dashed).  Host halo mass bins span $\Delta \log_{10} M = 0.5$,
centered on the mass value listed.  The upper left panel presents
rates measured at subhalo infall --- i.e., the merger rates of distinct
halos --- and the upper right panel presents rates of subhalo
destruction (when the associated subhalo loses $90\%$ of the mass it
had prior to entering the virial radius of the larger halo), which we
expect to more closely trace the galaxy merger rates.  The lower left
panel presents infall rates, now plotted at a fixed mass ratio ($>m/M
= 0.1, ..., 0.7$ from top to bottom) and host mass ($M = 10^{12.5}
\Msun$, triangles; $M=10^{11.5} \Msun$, squares) as a
function of redshift.  The same information is presented in the lower
right panel, but now presented as the rate per unit redshift instead
of per unit time.  We see that merger rates increase with increasing
mass and decreasing mass ratio, and that the merger rate per unit time
increases with increasing redshift out to $z \sim $4.

We quantify the measured dependencies using simple fitting functions.
The merger rate (for both infall and destruction rates) per unit time
for objects with mass ratios larger than $m/M$
into halos of mass $M$ at redshift $z$ is fit using
\begin{eqnarray}
\frac{dN}{dt} (>m/M) & =  & A_t(z, M) \, F(m/M).
\end{eqnarray}
For the infall rate, we find that the normalization evolves with halo mass and redshift as
$A_t(z,M) =0.02 \, {\rm Gyr}^{-1} \, (1+z)^{2.2} \, M_{12}^b$ with $M_{12}$ the mass in units of
$10^{12} \Msun$ and $b = 0.15$.  The merger mass ratio dependence is fit by
\begin{eqnarray}
 F(m/M) \equiv \left(\frac{M}{m}\right)^{c} \, \left(1 - \frac{m}{M}\right)^{d},
\end{eqnarray}
with $c = 0.5$, and $d = 1.3$.  A similar fit describes the destroyed
rate, as summarized in Table 1.
The fits are illustrated by solid lines that track the simulation points
in each of the $dN/dt$ panels in Figure 1.

The solid and dotted lines in the lower-right panel of Figure 1 show that
the infall rate per unit redshift, $dN/dz$,
is well described by the same mass-dependent function, but with a  normalization
that is only weakly dependent on redshift:
\begin{equation}
\frac{dN}{dz}(>m/M) = A_z(z,M) \, F(m/M),
\end{equation}
where $A_z(z,M) = 0.27  \,  (d\delta_c/dz)^2 \, M_{12}^{0.15}$ (for infall rate; see
Table 1 for destruction rate).
As discussed by \cite{FakhouriMa08} (hereafter FM08), a redshift evolution of this form is motivated by the
expectations of Extended Press-Schechter theory.  Note that since $d\delta_c/dz$ asymptotes
to a constant $\sim 1.3$ for $z \gtrsim 1$ and evolves only mildly to $\sim 0.9$ at $z \simeq 0$, the
overall redshift dependence is weak.

To a large extent, our results confirm and agree with those of FM08,
who studied merger rates for halos in the Millennium simulation
\citep{springel05} and presented a fitting function for the merger
rate per unit redshift per unit mass-ratio for halos as a function of
mass and redshift (the differential of our rate, $dN/dz$,
with respect to the merger rate $m/M$), and concluded that it was nearly universal in form.
For comparison, the pink dashed lines in the top left panel of
Figure 1 show the implied expectations based on the FM08 fit for
$M=10^{12} \Msun$ (lower lines for each pair) $M=10^{12} \Msun$ (upper lines)
halos.~\footnote{We use their fit for the `stitching'
merger rate, which corresponds most closely to our own definition
for halo mergers.}  The agreement is quite remarkable, especially in
light of the fact that the simulation, merger tree algorithm, and
halo finder all differed substantially from our own.  Note that the
agreement is particularly good over the mass ratios $m/M \simeq 0.05 -
0.5$, that are likely the most important for galaxy formation (in terms
of their potential for morphological transformation and overall mass
deposition, see \citealp{Stewart07}).
We note however that our infall rate data are smaller than FM08
by a factor of $\sim1.5$ for very large mass-ratio mergers $m/M
\gtrsim 0.7$ and by a factor of $\sim2$ for very small mass-ratio mergers $m/M
\lesssim 0.01$ (this discrepancy for small mergers is slightly worse at
low redshift, $z\lesssim0.3$).  In addition, we find a slightly stronger mass
dependence, $dN/dt \propto M^{0.15}$ as opposed to $dN/dt \propto
M^{0.1}$ as found by FM08.

It is interesting to note that in an independent analysis of the Millennium simulation,
\cite{Genel08b} studied the (infall) merger rates of halos by
defining halo masses and mergers in slightly different ways from FM08, in an effort to
further remove artifacts of the halo-finding algorithm of the simulation.
Among other results, their findings suggested that the merger rates from
FM08 are slightly too high (by $\lesssim50\%$) for low redshift and
for minor ($<1/10$) mergers.  This is qualitatively
similar to the differences between FM08 and our own results, motivating
the need for future study regarding the sensitivity of merger statistics from
dark matter simulation on halo finding algorithms, as well as halo mass and merger
definitions.

Our results also largely agree with an investigation of the major merger
rate ($>1/3$ mergers) of
halos and subhalos by \cite{Wetzel08}.  They found that
the infall rate for halos ($M=10^{11}-10^{13}\Msun$) evolves with redshift as $dN/dt =
A(1+z)^{\alpha}$ (with $A\sim0.03$ and $\alpha = 2.0-2.3$) from $z=0.6-5$,
in good agreement with our infall rates both in slope and in normalization (see Table
1).   \cite{Wetzel08} also reported on the subhalo merger rate in
their simulations (the rate at which satellite subhalos finally merge with the central
subhalo)
and found similar behavior as
field halos for low redshift ($A\sim.02$ and $\alpha\sim2.3$ for $z=0.6-1.6$,
in good agreement with our destruction rates)
but with a significantly flatter slope for high redshift
($A\sim0.08$ and $\alpha=1.1$ for $z=2.5-5$, a factor of $2-3$ lower than our results,
with a significantly flatter slope).
Even though
our destruction rate attempts to track a similar physical phenomenon
as their subhalo merger rate---the rate of impact of
satellite galaxies onto central galaxies---we find destruction rates to show
qualitatively similar behavior to infall rates at \emph{all} redshifts.

We speculate that the
discrepancy between their results and ours may be due primarily to differences in
definition.  For example, we define an infalling halo to be ``destroyed'' once it loses
$90\%$ of its infall mass (see \S 2).
\cite{Wetzel08} defines subhalo mergers by tracking the evolution of
the subhalo's $20$ most-bound particles,  resulting in a much
more stringent definition of a merger, and increasing the time delay between
infall to the virial radius ($t_{\rm inf}$) and the time at which
the satellite is destroyed ($t_{\rm merge}$).
More importantly, we define the merger mass ratio by the
halo masses when the satellite halo first falls into the virial radius of
the host, $m_{\rm inf}/M_{\rm inf}$.
Although we track the subhalo until it has lost $90\%$ of its mass in order to assign
a proper \emph{time} that the merger takes place, we do not redefine this merger ratio
based on any subsequent growth or decay of either halo.  Although \cite{Wetzel08} defines the
satellite halo's mass in an identical fashion, they allow for the growth of the central
halo during the decay time of the subhalo. Once the subhalo is destroyed,
they use the host halo mass at \emph{this} time (minus the mass of the subhalo, so that $m/M<1$)
and thus define the merger ratio as $m_{\rm inf}/M_{\rm merge}$.  As a consequence,
the host halo has a significant time period ($t_{\rm merge}-t_{\rm inf}$) to grow in mass,
leading to smaller mass ratio definitions for identical merger
events, as compared to our definition.  This effect is likely negligible at late times, when
halos do not grow significantly over the $\sim2$ Gyr decay timescales typical for major mergers.
This may be why the two studies agree rather well for $z<1.6$.
However, the central halo's mass growth on these timescales becomes increasingly important
at high redshift, possibly explaining the flattening of $\alpha$ reported by \cite{Wetzel08}, as
compared to our own results.

\cite{FakhouriMa09} investigates this issue to some degree by studying the subhalo merger rate in the Millennium simulation
using differing mass ratio definitions.
Whether they implement a merger ratio definition similar to
our destruction rate, or one more similar to that of \cite{Wetzel08}, their merger rates remain well-fit to a
power law in $(1+z)$, in line with our results.
In this case, the underlying cause of the discrepancy between our merger rates (and those of \cite{FakhouriMa09})
and the subhalo merger rates reported by \cite{Wetzel08} remains uncertain.
Such comparisons between our respective results highlight the differences that are manifest in
defining mergers.  When including baryons, properly defining mergers and merger mass ratios becomes even more complicated
\citep[see e.g.][]{Stewart09conf}, but even between dark matter structures, differences such as
those found between our work, \cite{Wetzel08} and \cite{FakhouriMa09} further motivate the need for focused
simulations in order to determine the timescales and observational consequences
associated with the much more cleanly defined rate with which dark matter
subhalos first fall within the virial radii of their hosts.

\begin{figure*}[htbp]
  \plotone{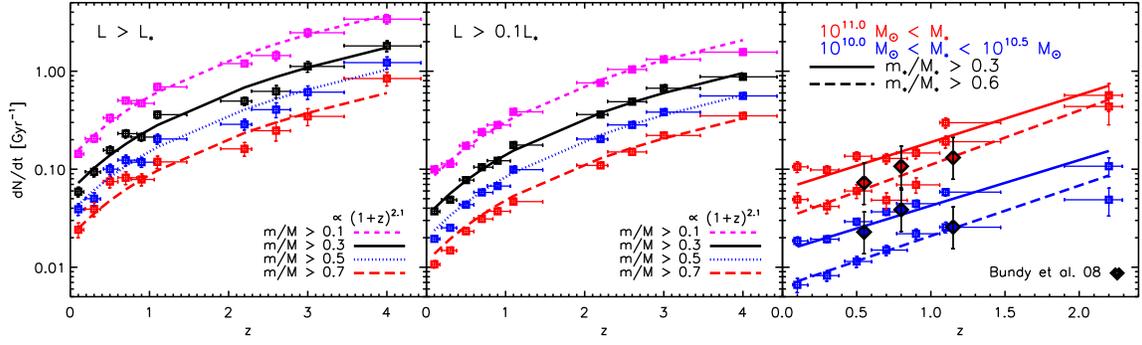}
  \caption{Expected merger rates per Gyr for galaxies of an indicated
    type as a function of redshift.  The vertical error bars show
    Poisson errors on both the number of main halos and the total
    number of mergers averaged over per redshift bin while the
    horizontal error bars show the redshift bins used to compute the
    merger rate at each redshift. The error bars do not include
    uncertainties in the mapping of mass to luminosity or stellar
    mass.  \emph{Left:} Merger rate into galaxies with $L > L_*$
    involving objects with {\em total} mass ratios $m/M > 0.1, ...,
    0.7$ as indicated.  \emph{Middle:} Merger rate into galaxies with
    $L>0.1L_*$.  \emph{Right:} Merger rates for galaxies of a given
    {\em stellar} mass involving objects with {\em stellar mass
      ratios} $m_*/M_* > 0.3$ and $0.6$ as a function of redshift. (Note that
    the redshift range in this panel only goes to $z=2$.)  We
    include two different stellar mass cuts in this panel, represented
    by the red and blue lines.  The dotted lines in this panel show an
    extrapolation out to $z\sim4$, based on our fit to the $z<2$
    simulation data.  The filled diamonds show observational results
    for the same stellar mass cuts from \cite{Bundy08}.}
\label{dNdt}
\end{figure*}

\section{Associating Halos with Galaxies}
\label{Methods}

While dark matter halo merger rates at a given mass are
theoretically robust quantities to compute in our simulation,
they are difficult to compare directly with observations.
One particularly simple, yet surprisingly successful
approach is to assume a monotonic mapping between dark matter halo mass $M$ (or similarly the halo
maximum circular velocity) and galaxy luminosity $L$
\citep{Kravtsov04a, Tasitsiomi04, ValeOstriker04, Conroy06, Berrier06, Purcell07,Marin08,ConroyWechsler08}.
With this assumption, provided that we know the cumulative number
density of galaxies brighter than a given luminosity, $n_g(>L)$, we may determine the associated halo population
by finding the mass $M(L)$ above which the number density of halos (including subhalos) matches that of
the galaxy population $n_h(>M_{\rm DM}) = n_g(>L)$.
Table \ref{LstarTable} shows the number densities of various galaxy populations from redshifts $z = 0.1 - 4$
obtained using a variety of surveys for galaxies brighter than $L = f\, L_*$, where $f = 0.1, 0.4$, and $1.0$.
We list the associated number-density matched minimum dark matter halo mass in each case, $M_{\rm DM}$, and we use
this association to identify halos with galaxies below.  For example, from the top left entry of this table,
we see that $n_h(>10^{11.2} \Msun) = n_g(>0.1L_*)$ at $z=0.1$.

One important point of caution is that the luminosity functions used to make these assignments at different
redshifts vary in rest-frame band, as indicated in Column 2.
Specifically, one concern might be that UV luminosity at low redshift is not strongly
correlated with dark matter halo mass, so assuming such a correlation for high
redshift galaxies is not valid.  Unlike their low redshift counterparts, however,
there is a strong correlation in high redshift ($z>2$) galaxies between star formation and
total baryonic mass, as well as a trend for more UV luminous galaxies to be more strongly
clustered, suggesting that connecting UV luminosity to halo mass at these redshifts
is a valid technique \citep[see discussion in][and references therein]{Conroy08}.
Encouragingly, as shown in the two
$z=3$ rows, the number density of $f \, L_*$ galaxies
from Sawicki \& Thompson (2006; rest-frame UV) and Shapley et al. (2001; rest-frame V)
are quite similar.

We also note that the data from these various sources will contain uncertainties in the
number counts of galaxies from, e.g.~cosmic variance.  For example, the COMBO17/DEEP2 data
fluctuates about a nearly constant value ($\sim0.02-0.03 h^{3}$Mpc$^{-3}$)
from $z=0.3-1.1$, suggesting a $\sim30\%$ uncertainty in these values.
We find that a $30\%$ error in the observed number density typically translates
into a similar $30\%$ error in the assigned minimum halo mass in our simulation.
Since dark matter halo merger rates are only weakly dependent on halo mass
($\propto M_{\rm DM}^{0.2}$),
this should result in only a $\sim10\%$ uncertainty in our merger rates.  Thus, the merger rates
we present here should be relatively robust to small errors in observational uncertainties.
For example, if we adopt minimum halo masses (regardless of redshift) of
$M_{\rm DM}=10^{11.2}, 10^{11.7}, 10^{12.3} \Msun$ as corresponding to
$>0.1, 0.4, 1.0 L_*$ galaxies, respectively, our resulting merger rates change by $<25\%$
(typically $5-15\%$).

A related approach is to use observationally-derived stellar mass functions
and to assume a monotonic relationship between halo mass and stellar mass $M_*$.
Though a monotonic relationship between total stellar mass and dark matter mass
avoids the issue of color band that arises in luminosity mapping,
we cannot use it explore merger rates as a function of stellar mass above $z\sim 2$ because
 the stellar mass function is poorly constrained beyond moderate redshifts.
For our analysis, we will adopt the relation advocated by  \citet[][hereafter CW09;
interpolated from the data shown in their Figure 2]{ConroyWechsler08}.
For example, CW09 find that the halo mass
 $M$ associated with stellar masses of $M_* = (1, 3, 10) \times 10^{10} M_\odot$
at $z=0, 1, 2$  are $M(z=0) \simeq (2.5, 7.0, 47 ) \times 10^{11} \Msun$;  $M(z=1) = (4.0, 9.6, 41) \times 10^{11}$;
and $M(z=2)=(2.1, 3.9, 10) \times 10^{12} \Msun$.
We note that because this mapping between stellar mass and halo mass is not well fit
by a constant ratio, $M_*=f M_{\rm DM}$, merger rates in terms of \emph{stellar} mass
ratios show qualitatively different evolution with redshift (see \S\ref{section:mergerrates}).
This is primarily because mergers of a fixed dark matter mass ratio do \emph{not}
typically correspond to the same stellar mass ratio \citep[see][]{Stewart09conf}.

Note that while the dark matter halo merger rates presented in \S \ref{DM} give robust
theoretical predictions, the merger rates we will present in terms of luminosity (or stellar mass)
are sensitive to these mappings between halo mass and $L$ (or $M_*$).  In addition, it is difficult to
perform a detailed investigation into the errors associated with these mappings, as there
are inherent uncertainties in the luminosity and stellar mass functions,
especially at $z>1$.  It is also possible that the monotonic mapping
between halo mass and $L$ (or $M_*$) may break down at $z>1$
(see discussion in CW09).
These uncertainties must be kept in mind when comparing our
predicted merger rates (in terms of $L$ or $M_*$) to observations, especially at high redshift.
Nevertheless, the halo masses we have associated with a given
relative brightness should be indicative.

\section{Galaxy Merger Predictions}
\label{Results}

\subsection{Merger Rates}
\label{section:mergerrates}
Our predicted merger rates (per galaxy, per Gyr) and their evolution with redshift,
averaged over $L > L_*$ and $L > 0.1 \, L_*$
galaxy populations, are illustrated in the left and middle panels of Figure 2.
Rates are presented for a few selected {\em dark matter halo} mass ratio cuts $m/M > 0.1$, $0.3$, $0.5$, and $0.7$.
Here, galaxy merger rates are defined using the \emph{destruction rate}, when the infalling
subhalo is identified as destroyed in the simulation (see \S2).  The solid lines correspond to a fit
in the form of Equation (1), with the normalization evolving as
$A_t(z,f) \propto (1+f) \, (1+z)^{2.1}$ for $L > f L_*$ galaxies.  The explicit best-fit parameters for the merger rate
as a function of luminosity cut are given in Table 1.

For comparison, the right panel in Figure 2 shows the predicted
evolution in the merger rates per galaxy for two bins of stellar mass,
according to the CW09 mapping described above: $10^{10.0} M_{\odot}<
M_* < 10^{10.5} M_{\odot}$ (lower, blue) and $M_* > 10^{11} M_{\odot}$
(upper, red).  Shown are merger rates for two choices
of {\em stellar mass} ratio mergers, $(m_*/M_*) > 0.3$,
$0.6$ (solid and dashed lines respectively).  The solid and dashed
lines correspond to fits to our simulation results in the form of
Equation (1), with $A_t(z) \propto e^{1.2 z}$.  The explicit best-fit
parameters for these two stellar mass bins (as well as an intermediate
bin, $10^{10.5} M_{\odot}< M_* < 10^{11.0} M_{\odot}$) can be found in
Table 1.  Table 1 also provides best fit parameters for the function
$F(r)$ (Equation 2) where now we associate the ratio $r$ with the
stellar mass ratio $r = m_*/M_*$.  Note that we only show our simulation points for
$z\lesssim2$ in this panel, due to uncertainties in the stellar mass function
at high redshift.

While the galaxy merger rate cannot be observed directly, it can be
inferred using a number of different techniques.  Mergers that are
about to occur may be forecast by counting galaxy close pairs, and
close pair fractions are often used as a proxy for the merger rate.
The filled diamonds in the right panel of Figure 2 are recent
merger-rate estimates from the pair count study of \cite{Bundy08}, for the same two stellar mass bins shown in the
simulations (blue for the lower mass bin, red for the upper mass bin).
\cite{Bundy08} have used the simulation results of \citet{kw08}
to derive merger rates from the observed pair fraction.  Overall, the
trends with mass and redshift are quite similar and this is
encouraging.  However, the \cite{Bundy08} results correspond to
mergers with stellar mass-ratios larger than $m_*/M_* \gtrsim 0.25$.
Our normalization is a factor of $\sim2$ too high compared to this, and only matches if we
use larger merger-ratios $m_*/M_* \gtrsim 0.5$.  It is possible that
this mismatch is associated with the difficulty in assigning merger
timescales to projected pairs (see, e.g. Berrier et al. 2006).
It may also be traced back to uncertainties in assigning stellar masses to
dark matter halo masses, however, since merger rates have relatively weak
dependence on halo mass, it would require increasing our assigned stellar masses
by a factor of $\sim3$ in order to account for this discrepancy solely by errors
in assigning stellar mass (such an increase in stellar mass would result in
unphysical baryonic content for dark matter halos:
e.g., $10^{12}\Msun$ halo containing $M_*>10^{11}\Msun$).

There are a number of other observational estimates of the merger
rate based on pair counts of galaxies \citep[e.g.,][]{Patton02, Lin04, Bell06a,
Kartaltepe07,Kampczyk07,deRavel08,Lin08,McIntosh08,Patton08,Ryan08}.
We choose to compare our results to \cite{Bundy08} as a recent
representative of such work, primarily because it is more straightforward for
us to compare to samples that are defined at a fixed stellar mass and stellar mass
ratio.  It is also difficult to compare to many different observational
results on the same figure self-consistently, because different groups adopt
slightly different cuts on stellar mass (or luminosity) and on mass ratios for pairs.
We note that if we were to extrapolate our best-fit curves to higher redshift than
our data ($z\sim4$), we find good agreement between our simulation data and the
merger rate estimates using CAS (concentration, asymmetry, clumpiness)
morphological classifications from \citet{Conselice03} for galaxies with
$M_* > 10^{10} \Msun$.  However, the mapping between stellar mass and halo mass
adopted from \cite{ConroyWechsler08} is only valid to $z=2$ (and most
robust for $z<1$), so extrapolating these fits to $z\sim4$ is only a first-order
check, and should not be considered a reliable prediction.

\subsection{Merger Fractions}

Another approach in measuring galaxy merger rates is to count  galaxies
that show observational signatures of past merging events such as
enhanced star formation, AGN activity, and morphological disturbances.
Unfortunately, the timescale over which
any individual signature will be observable is often extremely uncertain, and will depend on the total mass and
baryonic makeup of the galaxies involved as well as many
uncertain aspects of the
physics of galaxy formation  \citep[e.g.][]{Berrier06,cox07,lotz08b,kw08}.
In order to avoid these uncertainties,
we present results for merger fractions using several choices for lookback timescale here.

\begin{figure*}[htbp]
  \plotone{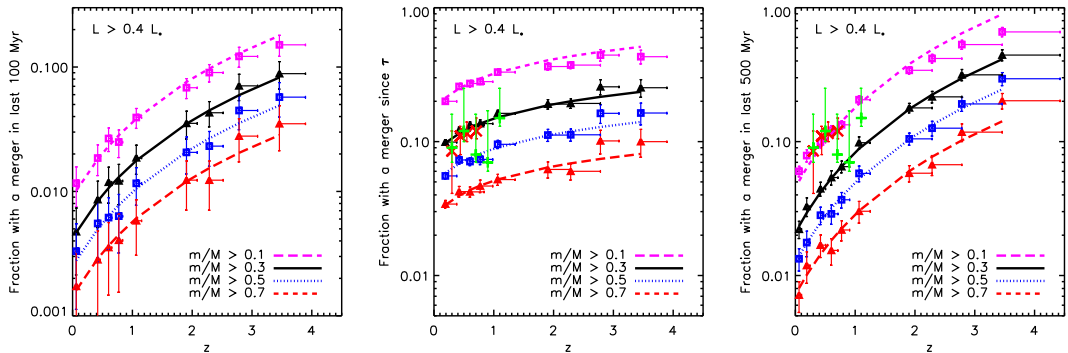}
  \caption{The fraction of halos that experience at least one merger
    larger than $m/M$ in the past 100 Myr (left), halo dynamical time
    $\tau$ (middle), or 500 Myr (right), as a function of $z$.  Error
    bars show the Poisson $\sqrt{N}$ error based on the both the
    number of main halos and the total number of mergers averaged
    over, while the horizontal error bars show the redshift bins used
    to compute the merger rate at each redshift.  The error bars do not
    include uncertainties in the mapping of mass to luminosity.  The
    symbols represent estimates of the observed merger
    fraction at various redshifts, based on \citet[][red crosses]{Jogee08} and
    \citet[][green plus signs]{lotz08a}, respectively.}
\label{Frac100}
\end{figure*}

The three panels of Figure \ref{Frac100} show the predicted evolution of the merger fraction in galaxies brighter than
$0.4 L_*$ for three different choices of merger lookback time and for various choices for the
total mass merger fraction $m/M > 0.1, ..., 0.7$.
The horizontal error bars on this figure show the \emph{actual} redshift bins used
to compute the merger fractions.
The left and right panels show the merger fraction within $100$ Myr and $500$ Myr,
respectively~\footnote{In most cases, the available timesteps ($\Delta t \simeq 250$ Myr)
are too widely spaced to directly measure fractions within 100 Myr.  For this reason,
 the left panel is actually the merger fraction within the last $\Delta t$ timestep,
scaled down by a factor of $(\Delta t/100 {\rm Myr}) \simeq 2.5$.},
 and the middle panel
shows the fraction of galaxies that have had a merger within the past halo dynamical
time~\footnote{We use $\tau = R/V \propto (\Delta_v(z) \, \rho_u(z))^{-1/2}$, such that
the halo dynamical time is independent of halo mass.}
$\tau(z)$, where
$\tau(z) \simeq 2.0$ Gyr $(1+z)^{-1.15}$ for $z \leq 1$ and $\tau(z) \simeq 2.6$ Gyr $(1+z)^{-1.5}$ for $z >1$.

\subsubsection{Merger-driven starbursts}

Several recent studies of star formation rates in galaxies at $z=0-1$
suggest that the cosmic SFR density is not dominated by strongly
disturbed systems with brief periods of intense star formation, as
might be expected if merger-driven starbursts are common.  Instead,
the SFR density appears to be dominated by normal, non-merging
galaxies \citep{Wolf05, Bell05, Jogee08, Noeske07}.
That is, $<30\%$ of the instantaneous SFR density at a given redshift (from $z=0-1$)
is derived from morphologically disturbed galaxies, which may be currently
undergoing a merger-induced starburst.  Even at high redshift ($z\sim2$),
a comparison of the clustering of star-forming galaxies to that
of dark matter halos suggests that these galaxies
are consistent with massive galaxies (in massive DM halos) quiescently forming stars,
as opposed to less massive galaxies (less massive DM halos) in the midst
of merger-induced starbursts \citep{Conroy08}.  However, this conclusion is based
on the assumption that UV-bright galaxies at this redshift comprise a
representative sample of star-forming galaxies.

As discussed in the introduction, the briefest timescales we expect for
merger-triggered starbursts is $\sim 100$ Myr
\citep{MihosHernquist96,cox07}, and for these models we expect the
SFR to increase to $\sim 20$ times the isolated value for
$m/M \gtrsim 0.3$ events \citep{cox07}.  (While we adopt these timescales as
``typical'' of galaxy mergers, it is important to keep in mind that
\cite{cox07} focuses on $z=0$ galaxies.  High redshift galaxies should typically
contain higher gas fractions, which may impact the properties
of merger-induced starbursts at these epochs.)  As we see from the left-panel
of Figure \ref{Frac100}, the fraction of galaxies that have a merger
large enough ($m/M > 0.3$) to trigger such a burst is quite small,
$\lesssim 1\%$ for $z \lesssim 1$.  It is therefore not surprising
that stochastic starbursts of this kind do not dominate the SFR
density at moderate to low redshifts.  Even at at higher redshift
($z=3-4$), the fraction of galaxies with major mergers on these
timescales is less than $ \sim 6\%$ of the total bright galaxy
population (consistent with the
results presented in \cite{Somerville08}, for their semi-analytic model).
However, galaxy gas fractions are expected to increase with redshift \citep{Erb06},
which could presumably result in significant starburst activity from more minor mergers
\citep[as well as providing fresh gas accretion in a more cumulative sense, see][]{Stewart09b}.
A higher fraction of galaxies have experienced such minor ($>1/10$) mergers on
these timescales at $z=3-4$ ($\sim15\%$).

Alternatively, if merger-driven starbursts remain active for $\sim
500$ Myr, as other models suggest, then their enhancements are
expected to be less pronounced (with an SFR $\sim 5$ times isolated;
\citealt{cox07}).  In this case, the right panel of Figure
\ref{Frac100} is the relevant prediction, and we see that (at most)
$\sim 3 - 9 \%$ of bright galaxies could exhibit signs of such
elevated SFR activity between $z=0$ and $z=1$.  It seems that in either
case, we would not expect merger-triggered activity to play a major
role in driving the integrated star formation rate at these epochs.
Only at the highest redshifts $z \gtrsim 3$ would this seem possible.
However, we once again point out that the detailed study
of \citealt{cox07}, which we have quoted here,
focuses on low redshift galaxies, with gas fractions $<30\%$.
If minor mergers with very high gas fractions ($>50\%$)
are capable of triggering starbursts, then over half of all bright galaxies
at $z>2$ (where such high gas fractions are more common) may
be in the process of starbursting.

Under the presumption that only major mergers trigger starbursts,
we note that our numbers are an upper limit on
the fraction of bright galaxies that could be experiencing merger-induced starbursts,
because moderately high gas fractions are also necessary.
For example, a study of $216$ galaxies at
$z\sim2-3$ by \cite{Law07a} found that
galaxy morphology (in rest-frame UV) was not necessarily correlated with star formation
rate, and in a recent examination of two Chandra
Deep Field South sources using adaptive optics, \cite{Melbourne05} found an example of
a merger of two evolved stellar populations, in which the major merger signature was not
accompanied by a burst of star formation, presumably because both galaxies were gas-poor.
\cite{Lin08} finds that $\sim8\%$ ($25\%$) of mergers at $z\sim1.1$ ($0.1$) are gas-poor,
suggesting that this issue, while less dominant at high redshift, is a significant effect
and must be taken into consideration.

We note that we have focused on galaxies which are in the \emph{midst} of
a merger-induced starburst.  The lingering impact these bursts will have on the
cumulative star formation histories (SFH) of galaxies in a separate issue entirely.
A recent study by \cite{CowieBarger08} traced recent star formation in $>2000$ galaxies from
$z=.05-1.5$ and found that roughly a quarter of these galaxies showed color indications
(AB3400-AB8140 vs. EW in H$\beta$) indicative of starbursts in the past $0.3-1.0$ Gyr.
Once again, we find our predictions to be broadly consistent with this result, with $\sim20\%$ of
bright galaxies ($>0.4L_*$) having experienced a major merger in the past Gyr at $z\sim1$.
With this study of individual galaxies' star formation histories emphasizing the
importance of starbursts, and the
previously mentioned studies of the global SFR density emphasizing the importance of star
formation in normal, non-merging systems, we find that our predicted merger rates are
broadly consistent with both results, suggesting that while starbursts may not be the
globally dominant form of star formation in the Universe, they still
play an important role in the star formation histories of galaxies.
Detailed progress in understanding the full importance of merger-induced starbursts on the
global SFR density of the Universe will require a better understanding of the
timescales and signatures associated with galaxy mergers and merger-induced starbursts.

\begin{figure*}[htbp]
  \plotone{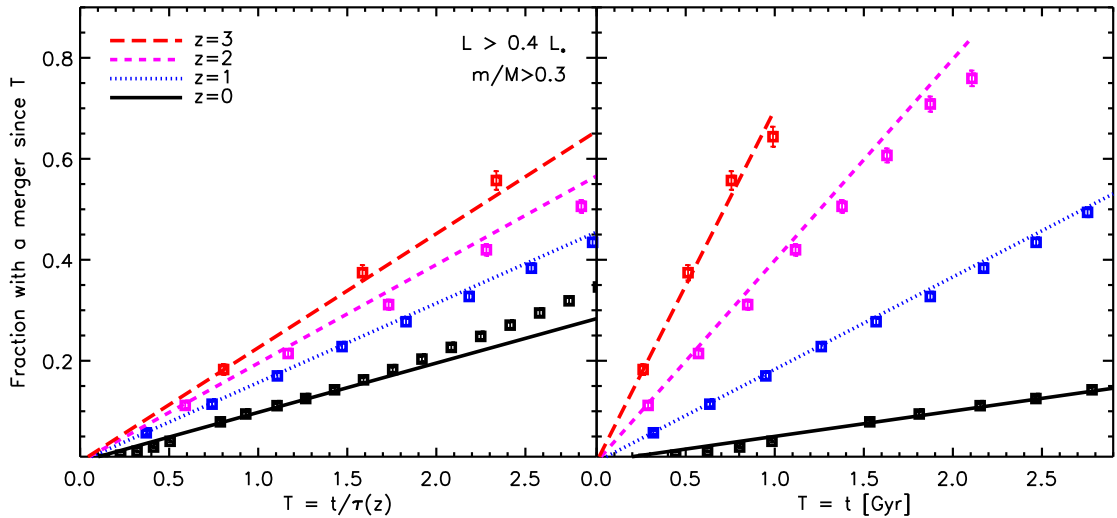}
  \caption{The fraction $>0.4L_*$ galaxies at $z \sim 0, 1, 2, 3$
    (black solid to red dashed lines) that have experienced a major
    merger ($m/M>0.3$) over a given time period.  Symbols show the
    simulation data, while the lines are given by the fit in Table 1.
    \emph{Left:} Merger fraction since $t$, normalized by the halo
    dynamical time at each redshift, $\tau(z=0,1,2,3)\simeq 1.95,
    0.92, 0.49, 0.32$ Gyr.  \emph{Right:} Merger fraction in the past
    $t$ Gyr.  Error bars show the Poisson $\sqrt{N}$ error based on
    the total number of mergers, and are comparable to the symbol
    sizes.  The error bars do not take uncertainties in the mapping of
    mass to luminosity into account.}
\label{Fracttau}
\end{figure*}

\subsubsection{Morphological signatures}
\label{Morphology}

Even if the contribution to the overall SFR due to very
recent mergers remains low, this does not necessarily imply that there would be
a lack of morphological signature.   The timescale for morphological relaxation
may be significantly longer than starburst activity.  Though the precise timescales
for relaxation are uncertain, the middle and right panels of Figure \ref{Frac100}
explore merger fractions for two reasonable choices: a fixed $500$ Myr timescale and
a redshift-dependent halo dynamical time $\tau$.

\cite{lotz08a} used AEGIS survey data to study the morphological evolution and implied galaxy
merger fraction from redshift $z=0.2$ to $1.2$.  The merger fraction results
for $>0.4L_*$ galaxies from \cite{lotz08a}
are shown by the green pluses in the middle and right panels of Figure 3.
In a similar investigation,  \cite{Jogee08}
study $z=0.2-0.8$ galaxies using a combination of HST, ACS, Combo-17,
and Spitzer 24 $\mu$m data to estimate the fraction of ``strongly disturbed'' galaxies.  Their
results~\footnote{The data from \cite{Jogee08} correspond to a fixed stellar mass cut
at $M_* \sim2.5 \times 10^{10} M_{\odot}$, but the associated dark matter halo mappings
from CW09 are close to those for galaxies with $>0.4L_*$ (see Table 1 and our discussion in \S 4).}
are shown by the red crosses in Figure 3, and are in reasonably good agreement
with the points from \cite{lotz08a}.
We note that the data from these very recent works
seem to be in good agreement with  the $m/M > 0.3$ merger fraction if the
relaxation time is close to $\tau$.  This case in particular has a fairly weak evolution because
$\tau$ is decreasing with time.  Interestingly,  however, due to the rather large measurement
uncertainties,  the data are also in reasonable agreement with the fixed relaxation timescale
case of $500$ Myr (which has a steeper evolution with $z$),
as long as more minor mergers ($m/M > 0.1$) can trigger the observed
activity.
The fact that the data matches both the predictions in the middle
panel and right panel of Figure 3 draws attention to the inherent degeneracies in
this comparison.  The same merger fractions are obtained with high-mass ratio
merger events and lookback times or with lower mass ratio mergers with slightly
shorter lookback times.

We may also compare our predictions with the results of
 \cite{Melbourne08},  who imaged 15 $z\sim0.8$ luminous infrared galaxies (LIRGs) with the
Keck Laser Guide Star (LGS) AO facility, and found that $3/15$ of the galaxies showed
evidence for a \emph{minor} merger, while only $1/15$ was consistent with a major merger.
These results match our expectations for major ($m/M > 0.3$) and minor ($m/M > 0.1$)
merger fractions at $z \sim 0.8$ fairly well, considering the
small number statistics.  Similarly, \cite{Shapiro08}
study 11 rest-frame UV/optical-selected z$\sim$2 galaxies with spectroscopic data
from SINFONI on the VLT, and estimate that $\sim$25$\%$ of these systems are likely undergoing
a major (mass ratio $\leq$ 3:1) merger.    Again, our expectations as shown in the middle
and right panels of Figure 3 are consistent with these numbers.

The above discussion makes it clear that meaningful comparisons
between observed morphologically disturbed fractions and predicted
merger fractions rely fundamentally on understanding how the mass
ratio involved affects the morphological indicator and on the
associated relaxation timescales of the associated remnants.  In
addition, merger rates are expected to depend sensitively on the
galaxy luminosity and redshift (see Table 1).  Comparisons between
observational results and theory therefore require great care,
especially as it concerns the evolution of the merger rate.   If, for
example, higher redshift measurements are biased to contain brighter
galaxies than lower redshift measurements, then the redshift evolution
will likely be steeper than the underling halo merger rate at fixed
mass.  Or, if higher redshift measurements are sensitive to only the
most massive mergers, while lower redshift measurements detect more
subtle effects, then the evolution in the merger rate will be biased
accordingly.

\subsubsection{High Redshift Expectations}

As seen clearly in Figures 2 and 3, the merger rate per galaxy and the corresponding
merger fraction at a fixed time are expected to rise steadily towards high redshift.
Even after normalizing by the halo dynamical time, which decreases with redshift,
this evolution with redshift persists, as seen in Figure 3 (middle).
This point is emphasized in Figure 4, which shows the fraction of
$L>0.4 L_*$ galaxies that have had a merger larger than $m/M=0.3$ within
the last $t$ Gyr (right) and within the last $t/\tau(z)$ (left).
The left-hand panel scales out the evolution in the halo dynamical
time.  We see that $\sim50\%$ of z=3 galaxies are expected to have had a
major merger in the last 700 Myr, and that these galaxies are $\sim4$ times
as likely to have had a significant merger in the last dynamical time
than bright galaxies at z=0.  It would be surprising then if mergers
did not play an important role in setting the the properties of most
$z=3$ galaxies like Lyman Break Galaxies (LBGs).  These major mergers
should (at least) deliver a significant amount of gas to fuel star
formation, affect LBG dynamics, and perhaps trigger starburst activity.
If LBGs represent a biased sample at $z = 3$ (of unusually bright
galaxies, more likely to have recently undergone a merger-induced starburst)~\footnote{It
is estimated that $\sim 75\%$ of all bright galaxies at $z \sim 3$
are LBGs \citep{Marchesini07, Quadri07}.}
then it may be possible that the merger fraction in LBGs is even higher
than the global merger fraction for $>0.4L_*$ galaxies.

At higher redshifts, $z >3$, we expect major mergers to become
increasingly common.  The brightest galaxies $L > 0.4 L_*$ should be
undergoing mergers frequently, with an overwhelming majority of $z=4$
galaxies having experienced some significant merger activity in the
last $\sim 500$ Myr.

\section{Conclusion}
\label{Conclusion}

We have used a high-resolution $\Lambda$CDM $N$-body simulation to
investigate the instantaneous merger rate of dark matter halos as a
function of redshift (from $z=0-4$), merger mass ratio, and host
halo mass from $M = 10^{11}$ to $10^{13}
\Msun$.  Merging companions as small as $m = 10^{10} \Msun$ were
tracked.  We use number density matching to associate galaxies with
dark matter halos and present predictions for the merger rate and
merger fraction as a function of galaxy luminosity and stellar mass.  The
principle goal has been to present raw merger statistics that can be
compared directly to observations of galaxies to high redshift.
Fitting functions that describe our results as a function of
luminosity, mass, mass-ratio, and redshift are provided in Table 1.

Our main results may be summarized as follows:

\begin{enumerate}

\item A simple fitting function describes the accretion rate of small
  dark matter halos of mass $m$ into larger dark matter halos of mass
  $M$ as a function redshift: $dN/dt = A(z,M)\, F(m/M)$, where typically
  $A(z,M) \propto (1+z)^{2.2} M^{0.15}$ and $F(m/M)=(M/m)^c (1-m/M)^d$.
  Fit parameters for merger rates
  in terms of dark halo mass, luminosity, or stellar mass are given in Table 1.

\item The merger rate of galaxies of luminosities $L>f L_*$ should
  evolve in a similar manner, with a redshift and luminosity
  dependence that follows $A(z,f) \propto (1+f) \, (1+z)^{2.1}$.

\item Only a small fraction ($0.5\%$ at $z=0$, $10\%$ at $z=4$) of
  bright ($> 0.4 L_*$) galaxies should have experienced a major ($> 0.3$)
  merger in their very recent history (100 Myr, Figure 3 left panel).
  Even if mergers trigger the kind of short-lived, highly-efficient
  star formation bursts that are expected in some models, they cannot
  contribute significantly to the overall distribution of star
  formation rates at any given epoch.

\item The predicted fraction of galaxies with a merger in the past 500
  Myr, or alternatively within a past halo dynamical time, are in
  reasonable agreement with the fraction of galaxies that show
  observational signs of morphological disturbance between redshifts
  $z = 0 - 2$ (Figure 3, middle and left panels).  We emphasize, however, that
  comparisons between theory and observations suffer from significant
  uncertainties associated with mass-ratio dependencies and relaxation
  timescales.

\item Galaxy merger rates should depend on at least three parameters:
  mass (or luminosity), merger mass ratio, and redshift  (see Table 1).
  Therefore any attempt to compare two observational indicators of the
  merger rate or to relate specific observations to
  theoretical predictions must take great care in the respective comparisons.

\item Mergers must become increasingly important in shaping galaxy
  properties at $z>3$.  At $z=3$, the fraction of galaxies with a
  merger in the past dynamical time is $\sim 4$ times higher than at
  $z=0$.  We expect $\sim 30\%$ ($60\%$) of $> 0.4 L_*$ galaxies to
  have experienced a $m/M > 0.3$ major ($m/M > 0.1$ minor) merger in
  the past 500 Myr at $z=3$.  Though it is unlikely that short-lived
  starbursts associated with these mergers drive the increase in the
  global star formation rate of galaxies with redshift, the broader
  implications of these mergers (fresh supply of gas brought in to the
  central galaxy through accreted satellites, etc.) are undoubtedly
  linked to star formation and the general growth of galaxies on
  longer timescales.

\end{enumerate}

\acknowledgements The simulation used in this paper was run on the
Columbia machine at NASA Ames.  We would like to thank Anatoly Klypin
for running the simulation and making it available to us.  We are also
indebted to Brandon Allgood for providing the merger trees.  We thank
Charlie Conroy for providing us the abundance matching data from CW09,
and Kevin Bundy for providing us with an advance copy of his paper before publication.
We thank Jeff Cooke, David Law, Lihwai Lin, Ari Maller, David Patton, Brant Robertson, and
Andrew Wetzel for useful discussions.  We also thank the anonymous referee,
whose insightful comments helped us improve the quality of this paper.
JSB and KRS are supported by NSF grant AST 05-07916.  KRS, JSB, and EJB
received additional support from the Center for Cosmology at the
University of California, Irvine.  RHW was supported in part by the
U.S. Department of Energy under contract number DE-AC02-76SF00515 and
by a Terman Fellowship from Stanford University.

\bibliography{ms}

\end{document}